\documentclass[final,5p,times]{elsarticle}

\usepackage[linktocpage=true,unicode=true,ps2pdf]{hyperref}
\usepackage{hypcap}
\usepackage{amsmath,amssymb,amsfonts}
\usepackage{graphicx}
\usepackage{color}

\graphicspath{{./figs/}}
\begin{document}

\begin{frontmatter}

\title{ Inconsistence of super-luminal Cern-Opera neutrino speed with observed \\
    SN1987A burst and neutrino mixing for any imaginary neutrino mass}
\author{Daniele Fargion, Daniele D'Armiento}
\ead{daniele.fargion@roma1.infn.it}
\address{Physics Department,  Rome University 1, Sapienza
and INFN, Roma1 --Pl. A. Moro 2, 00185, Rome, Italy.}

\begin{abstract}We tried  to fit in any way the recent Opera-Cern claims of a neutrino super-luminal speed with observed Supernova SN1987A neutrino burst  and all (or most)  neutrino flavor oscillation.
   We considered three main frame-works: (1) A tachyon imaginary neutrino mass, whose timing is nevertheless in  conflict with observed IMB-Kamiokande SN1987A burst by thousands of billion  times longer. (2) An ad hoc anti-tachyon model whose timing  shrinkage may accommodate SN1987A burst but greatly disagree with energy independent Cern-Opera super-luminal speed. (3) A split  neutrino flavor speed (among a common real mass relativistic $\nu_{e}$ component and a super-luminal $\nu_{\mu}$) in an ad hoc frozen speed scenario that is leading to the prompt neutrino de-coherence and the rapid  flavor mixing (between $\nu_{e}$ and $\nu_{\mu}$, $\nu_{\tau}$) that are in conflict with most oscillation records.   Therefore we concluded that an error must be hidden  in Opera-Cern time calibration (as indeed recent rumors seem to confirm). We concluded reminding the relevance of the real guaranteed  minimal atmospheric neutrino mass whose detection may be achieved  by a milliseconds graviton-neutrino split time delay among gravity burst and neutronization neutrino peak in any future  SN explosion in Andromeda recordable in Megaton neutrino detector.
\end{abstract}


\end{frontmatter}

\section{Introduction: Any solution for super-luminal neutrinos?}
A first preprint from Cern-Opera experiment  hint for a muon neutrino faster than light \cite{Opera1}, may be tachyon in nature. If all neutrino were just tachyon their arrival (at SN1987A-17 MeV energy) would be even much much faster than a 17 GeV Opera neutrino. Indeed Opera super-luminal neutrinos (at a speed $2.5 \cdot 10^{-5}$ times faster than c), would lead to a SN1987A speed nearly 6.95 times faster than c, coming therefore much earlier, back nearly 134500 years ago from Large Magellanic Cloud, therefore unobservable. On the other side if  all the neutrino velocity, independently on their energy, were frozen at a Opera speed $2.5 \cdot 10^{-5}$ times faster than c,  than Supernova 1987A had not to be observed (as it is well known to be) on February 23th 1987, but just  3.72 years before, in late 1982 early 1983: their signals would be eventually hidden in oldest IMB records. However in such tuned new physics no explanation will be of the same neutrino burst found on February 23 1987 by IMB-Kamiokande. An ad-hoc anti-tachyon neutrino law (opposite energy relation respect tachyon) may somehow fit the super-luminal result and SN87A but it disagrees with apparent energy independence in Opera $\nu$ speed.   A more accommodating scenario is the one where electron neutrinos (and antineutrino) fly near velocity c, while muon neutrino are super-luminal: than SN1987A $\nu_{e}$ $\bar{\nu_{e}}$ may be in agreement with observed signals; nevertheless even in this ideal scenario  one should also find a coexisting precursor neutrino  burst signal in early 1982-1983 inside IMB records (certainly unobserved) , signal due to a partial  muon to electron neutrino conversion in flight from the SN1987A  to Earth. Moreover the electron muon different velocity is in obvious conflict with flavor interferences. Any different $\nu_{e}$ $\bar{\nu_{e}}$ speed respect $\nu_{\mu}$ $\bar{\nu_{\mu}}$  strongly disagree with all the observed oscillations as the near distance neutrino flavor  mixing in atmospheric neutrino (either muonic and in particular of electronic flavor in Super Kamiokande) as well as  in   Kamland  electron neutrino oscillation record.  Even Opera and Minos muon neutrino flux should have had to suffer by a prompt super-luminal  muon neutrino de-coherence from slower electron $\nu$ flavor in flight. In conclusion observed  SN 1987A neutrino burst and known neutrino mixing strongly constrain any ad hoc super luminal neutrino signal. Apparent Opera anomalous  neutrino speed measure might be indebt, we claimed, to some miss-leading time calibrations. Of course we didn't comment here the long list of puzzle in such violating special relativity, where one may imagine to sit along the neutrino super-luminal frame seing inverted time sequence of events. Surprisingly very recent test and preprint with unique sharp bunches  from CERN  once again reconfirmed such unbelievable (but widely applauded) super-luminal result \cite{Opera2}. We didn't change our mind.  However last minute rumors  of  experimental OPERA bugs finally shut down these, let say, imaginary results \cite{Rumors}. Nevertheless future Supernova gravitational waves  a  millisecond  time precursors  (respect neutrino burst due to SN neutronization) from Andromeda  may finally discover  neutrino mass splitting, mostly of real guaranteed atmospheric nature.
\section{Time precursor for imaginary tachyon}

Let us assume, as Opera-CERN declared, that the time precursor neutrino arrival is ${\delta(t)_{\nu}}= 60$ nanosecond.
Its velocity of light fly-time on 720 km distance is $\delta(t)_{Cern-Opera}= 2.4 $ ms. It implies for an energy independent
neutrino speed nature, a precursor event at a-dimensional time

\begin{equation}
{\frac{\delta(t)_{\nu}}{\delta(t)_{Cern-Opera}}} \simeq 2.37 \pm 0.32 \cdot 10^{-5}
\end{equation}
and a consequent \emph{apparent} precursor explosion from a SN1987A would be occurred 3.72 years before (the 23th February 1987)  optical SN event reaching from 157k ly (light year) distances in Large Magellanic Cloud.
 Probably around 2th June 1983 (incidentally on Italian Nation Day).
 But this result do not takes into account of the needed  tachyon neutrino  behavior, where the energy is related to an imaginary mass time by a Lorentz factor $E_{\nu_{\mu}} =  i m c^{2}\cdot {\gamma_{\nu_{\mu}}}$.
The Lorentz factor ${\gamma_{\nu_{\mu}}}= \frac{1}{\sqrt{1-(\beta_{\nu_{\mu}}})^{2}}$ for a super-luminal particle is an imaginary value.
Indeed the higher the energy (Opera 17 GeV) the slower (nearer to velocity of light) the  speed. The lower the neutrino energy the faster its speed; in this case SN neutrino is nearly 6.95 time faster than c:
 $\beta_{SN}= \sqrt{(\beta_{OPERA}^{2}-1)\cdot (\frac{E_{Opera}^2}{E_{SN}^2}+1) }$.
 The time arrival for a lower energy (let say 17 MeV SN1987A neutrino) SN precursor one should be nearly $134500$ years ago, assuming a LMC distance of 157k ly (light year). A disagreement of nearly a thousand of billion times the observed SN 1987A neutrino time scale.

\subsection{An anti-tachyon to save Opera and SN1987A $\nu$ timing}
 Let us just try for a while to fit this wrong SN1987A timing,  imposing, just for hypothesis, a \emph{an invented ad hoc tachyon-like relativistic law}, opposite to usual one :$E_{\nu} = - i m c^{2}\frac{1}{\gamma}$  with same expression for all flavor neutrinos, but whose different masses allow flavor mixing,  just \emph{almost} able to fit the Opera observation and the SN1987A burst signal. This law may have a minimal physical
 connection (respect to the above  tachyon law) if one assumes that the new tachyon  neutrino effective  mass $\hat{m_{\nu}}$ does depends on its  speed in matter as $\hat{m_{\nu}} = -  m \frac{1}{\gamma^{2}}$  ; one than obtains

$$
{\frac{\delta(t)_{\nu}}{\delta(t)_{Cern-Opera}}} \simeq 2.5 \cdot 10^{-11}
$$
   Therefore SN neutrinos fly  almost at light velocity. This time spread corresponds nevertheless to a two minutes
   spread for the supernovas 1987A neutrino arrival from Large Magellanic Cloud. A value barely consistent with Kamiokande records and the IMB one signal spread:  twelve  sec. Just comparable in global time, but not in details. Assuming an even more ad hoc law ($E_{\nu} = - i m c^{2}\frac{1}{\gamma^{1.166}}$) one may reconcile the time spread within 12 s.
    However  both these new ad hoc tachyon laws strongly disagree  with the negligible spread in different energies of the neutrino speed observed in OPERA itself: at a nominal Opera neutrino energy of 13.9 GeV the neutrino arrival is 53.1 ns earlier than c, while at 42.9 GeV the arrival is a little earlier, 67.1 ns before c; an observed difference of nearly $21\%$. On the contrary the $E_{\nu} = - i m c^{2}\frac{1}{\gamma}$  law would require at those higher energies (scaled by a factor 3.1 respect lower ones)  an earlier arrival of neutrino $3.1 ^{2}$ earlier, about 477 ns., or at a time difference above $900\%$ the lower energy ones.  Therefore the new tachyon law adapted to solve the SN1987A is in conflict with the OPERA almost un-variability of the neutrino speed with the energy.
      In conclusion this simplest anti-tachyon  toy model has some global fit, but it is extremely unnatural and nevertheless inaccurate and against OPERA  neutrino speed at two different energy. The extension to fit also the mixing among flavors is not forbidden but call for unnatural fine tuned tachyon masses values. Indeed the anti-tachyon mass value for the $E_{\nu} = - i m c^{2}\frac{1}{\gamma}$ law in Opera  requires $2.4$ TeV energy calling to a thousand billion time tuned mass splitting to solve observed flavor neutrino mixing. Therefore, because of all these failure, we try to accommodate OPERA result assuming, as a last attempt, that the muon (OPERA) and electron (SN1987A) neutrino velocity behavior  is different and therefore uncorrelated.

       \begin{figure}[!t]
  \vspace{5mm}
  \centering
  \includegraphics[width=3.5 in]{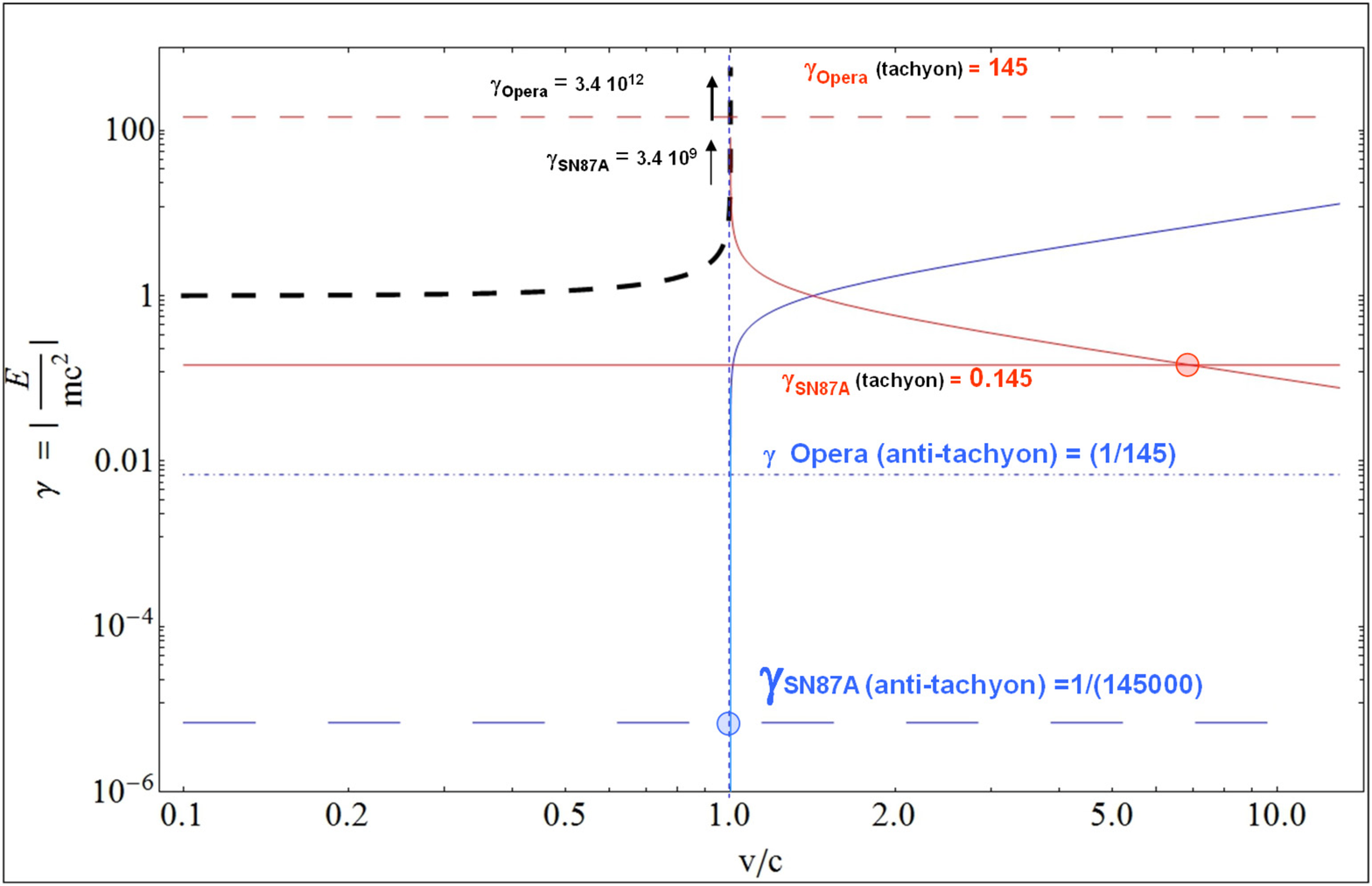}
  \caption{This is a schematic Energy-velocity, or better to say, Lorentz factor-velocity behavior for real neutrino ($v< c$) on left side (atmospheric neutrino mass), tachyon mass ($v> c$) right side red curve decreasing, anti-tachyon ($v> c$) right side blue curve growing, that are trying to fit at once Opera and  the neutrino SN1987A timing, correlated to the OPERA-Cern claim. We assume  OPERA neutrino at 17 GeV and SN1987A at 17 MeV.  Anti-tachyon described in figure  would shrinkage the timing almost as the observed ones.  Anti-Tachyon blue curve, whose  $E_{\nu} = - i m c^{2}\frac{1}{\gamma}$, requires a SN scale time spread nearly ten times longer the observed one. Assuming a rare  ad hoc  Anti-Tachyon  law ($E_{\nu} = - i m c^{2}\frac{1}{\gamma^{1.166}}$), one might tune energy-arrival dependence for OPERA and SN event, but Opera energy-speed spread should have been showing an (unobserved) strong velocity-energy dependence, nearly a factor  $900\%$ for the lower energy ones respect higher energy events. }
  \label{fig2}
 \end{figure}


\section{Frozen neutrino speeds: Looking back in 1983 IMB}
  Let assume, following also most recent 2011 TAUP conference, MINOS result, that there is no (much) differences between the observed  SN1987A $\bar{\nu_e}$ and the conjugate $\nu_e$.  In other words let us assume that we don't face any relevant CPT violation. Moreover let us  assume a frozen super-luminal neutrino velocity  (not energy dependent), only
for $\nu_{\mu}$,$\bar{\nu_{\mu}}$ flavors, as early CERN-OPERA result seem to favor \cite{Opera1}. In this scenario, if also electron neutrino share a frozen speed,  as we already wrote in the abstract,  there will be no room for any SN neutrino signal on 1987A: any burst of few second  would be too much hidden in a precursor event few years ( 3.72) earlier. If one really want to let survive SN1987A records with OPERA, he may call for (an unnatural) different flavor neutrino speeds: a scenario  where electron neutrinos (and antineutrino) fly nearly at  velocity c, while muon neutrino  $\nu_{\mu}$,$\bar{\nu_{\mu}}$ (as well as its mixed flavor  $\nu_{\tau}$,$\bar{\nu_{\tau}}$ in order to guarantee the solid $\nu_{\mu}$ $\nu_{\tau}$ mixing)  are super-luminal: than SN1987A $\nu_{e}$ $\bar{\nu_{e}}$ may be in agreement with observed signals; nevertheless even in this ideal scenario where  $\nu_{e}$,$\bar{\nu_{e}}$ are reaching in time the SN1987A optical burst, one should expect  a coexisting precursor neutrino  burst signal in late 1982  or early 1983 (just 3.72 years earlier)  inside IMB records.This because the $\theta_{12}$mixing angle coupling electron and muon flavors.  Kamiokande was born on late 1983 and cannot be searched in.  This SN1987A neutrino burst precursor presence  should rise in IMB detector because thermal SN1987A muon neutrinos will fly faster but their faster mass eigen-states should  also oscillate reaching the Earth as electron flavor $\nu_{e}$ $\bar{\nu_{e}}$. The same for the tau neutrinos whose presence maybe coeval with muon ones leading to a signal $3.72$ years earlier.  To find such a $\simeq 8$ neutrino event (or even $\simeq 16$ because eventual thermal tau neutrino conversion into electron ones) cluster in IMB will be, in my eyes, the real surprising revolution offered by OPERA. However nevertheless, any large different $\nu_{e}$ $\bar{\nu_{e}}$ speed respect $\nu_{\mu}$ $\bar{\nu_{\mu}}$ strongly disagree also with  other observed signal at low (MeV) and high (GeV) energy neutrino flavor  mixing, mostly the Kamland results, see Fig. \ref{fig6}, as well as the  correlated atmospheric electron and muon neutrino angular spectra  see Fig.\ref{fig3}. In such a model one would expect not only a muon neutrino anomaly in up-going vertical muon, but also a more dramatic  upward and downward electron neutrino suppression, due to the flavor de-coherence to be discussed below, effect that was never observed. In conclusion SN 1987A and known neutrino flavor mixing strongly disagree with any ad hoc super luminal neutrino model or with the present frozen muon neutrino super-luminal behavior.

 Anyway, without prejudice,  one may (or must) search in  oldest IMB records for the presence of any precursor twin neutrino burst in earliest  3.72 years since 1987, let say around June 1983, centered (within a spread of a couple of months) around 2 June 1983. The IMB detector was already recording since 1982 year, Kamiokande was not yet active. The presence of such a precursor (that for different reasons is unrealistic) will be  boosting the hypothetical  imaginary Opera-CERN discover from its present unacceptable  field to a more consistent experimental arena. An even more revolutionary discover may come from an additional twin cluster of event due (for instance) to a tau neutrino slightly different speed component; this possibility (additional split in muon versus tau neutrino velocities) is nevertheless  much unexpected  in view of the short oscillation scale well observed for muon neutrino conversion into tau ones by atmospheric SK  muon neutrino and also in K2K records. Indeed all such a frozen neutrino speed model should overcome many other test, basically all the observed mixing data, with very little hope of survival.

\section{Neutrino $\nu_{e}$ versus  $\nu_{\mu}$ in fast de-coherence}

\begin{figure}[!t]
  \vspace{5mm}
  \centering
  \includegraphics[width=3.3 in]{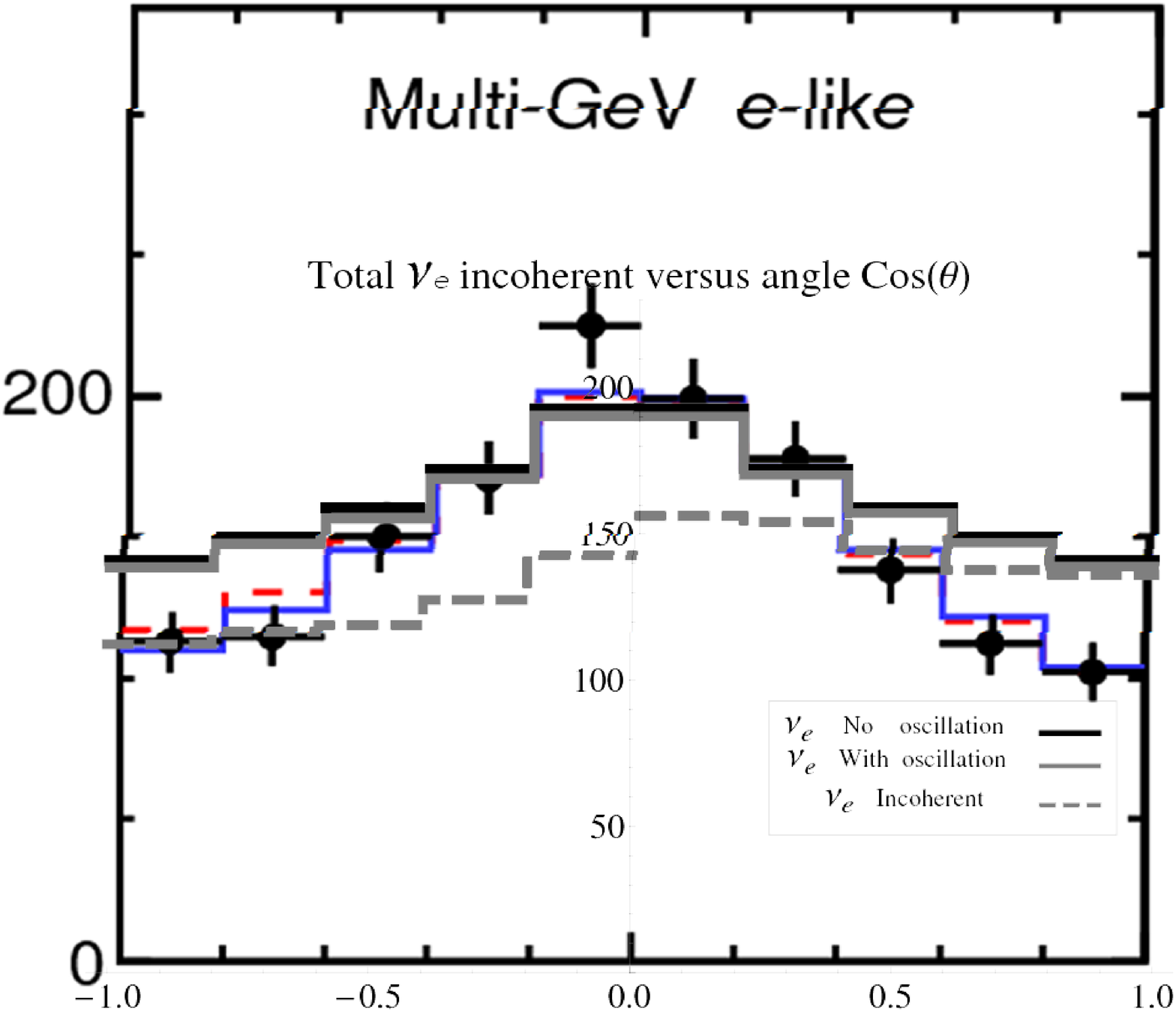}
   \includegraphics[width=3.3 in]{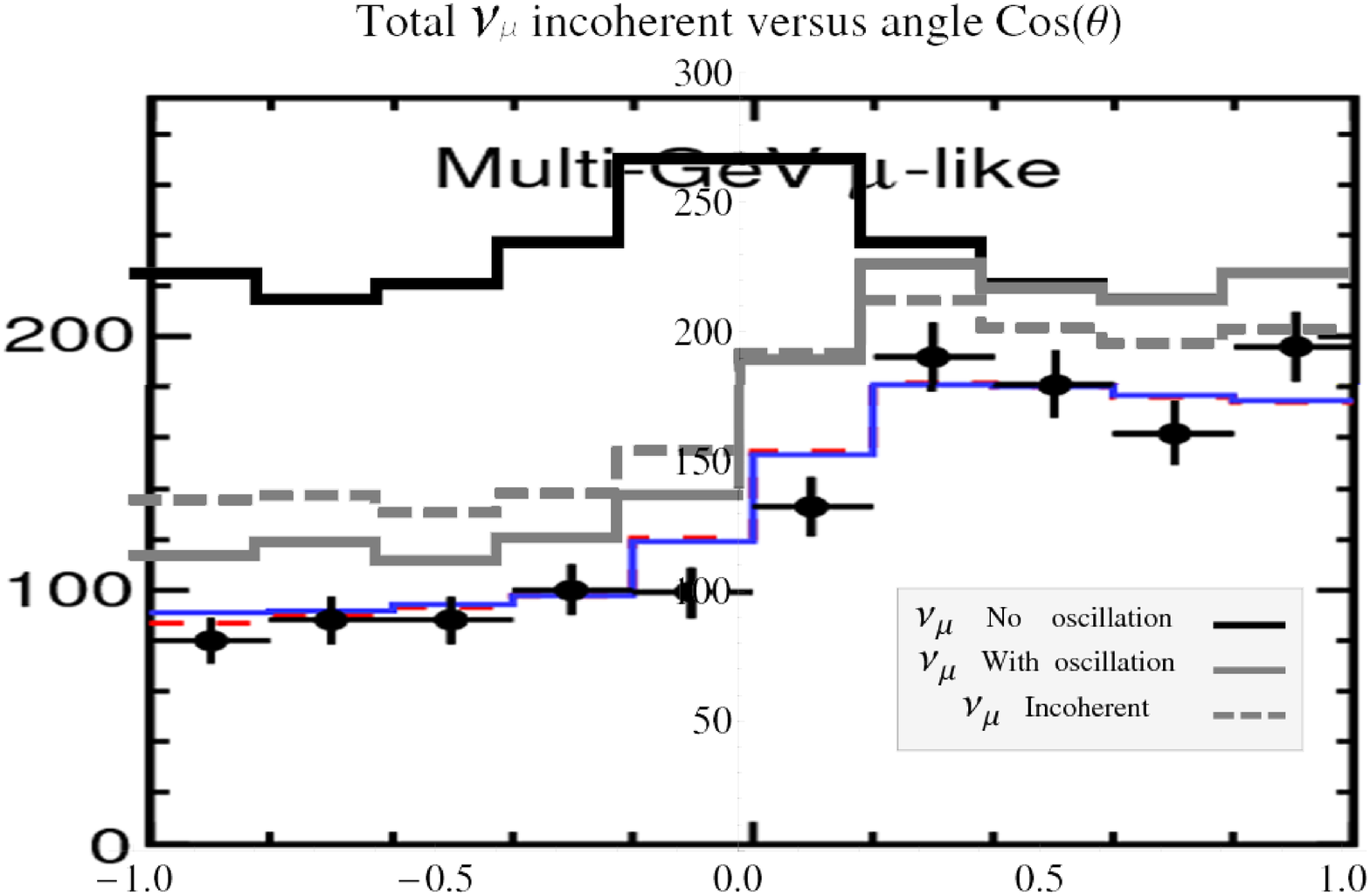}
  \caption{Our simulation of the expected zenith angle ($\cos\theta$) event count distribution for electron-muon $\nu$ mixing in the  atmospheric $\nu_{e}$ signals in top figure, $\nu_{\mu}$ signals in bottom figure, within a muon neutrino super-luminal scenario, where muon and electron $\nu$ are separated in speed and in fast de-coherence. The image is superimposed on last SK I,II, III data (2010)  on neutrino mixing. Time integral is 1489  days as in SK in PDG 2010. The energies range in the windows $1.33-10$ GeV.   The zenith angle distributions for fully contained 1-ring,  e-like
and only $\mu$-like events both with visible energy  $ > 1.33$ GeV. The back-ground black continuous histogram show the non-oscillating
Monte Carlo events, and the solid thick gray histograms show the best-fit expectations for common
neutrino oscillations \cite{Ka}. Our frozen speed electron-muon neutrino mixing is described by a dashed gray histogram made by a complex combination of
 effects as the muon (over electron) flux ratio, the different muon neutrino (over electron) cross section, the muon into tau oscillation at different zenith-distance tracks, and the different distances for pions and muons in decay in flight at each zenith angles. Dashed gray histogram describes  this new de-coherent scenario able to segregate the electron-muon flavor. However the departure from the  data ( mainly for electron flavors) is remarkable and in severe conflict with the observations. No way for frozen super-luminal neutrino speeds.}
  \label{fig3}
 \end{figure}

\begin{figure}[!t]
  \vspace{5mm}
  \centering
  \includegraphics[width=3.0 in]{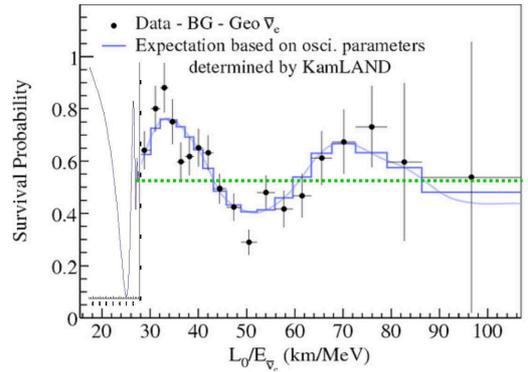}
  \caption{The expect the electron mixing and de-coherence in   Kamland neutrino signals (left side, inner box) in a super-luminal scenario, where muon and electron are separated in speed. The very fast de-coherence among electron and muon flavor states (here emphasized and expanded in size) occurs within micron meter distances. It may suppress the primary nuclear $\bar{\nu_{e}}$ in a sharp (0.04 $\mu$ m) distance by a large factor , $P(\nu_{e}\rightarrow \nu_{e}) = 0.547$, as shown by dotted horizontal line, a steady signature far from the observed  oscillating one in Kamland data}
  \label{fig6}
 \end{figure}

\begin{figure}[!t]
  \vspace{5mm}
  \centering
  \includegraphics[width=3.0 in]{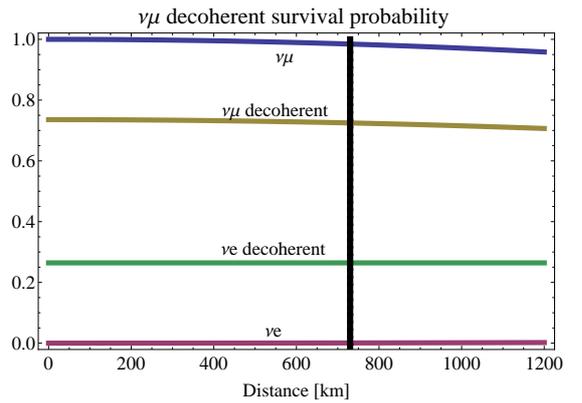}
  \caption{The expect muon,  electron and tau mixing  due to muon electron neutrino de-coherence in   OPERA experiment. Note the suppression due to the probability  $P(\nu_{\mu}\rightarrow \nu_{e})=P(\nu_{e}\rightarrow \nu_{\mu}) = 0.264$}
  \label{fig7}
 \end{figure}

Once again, assuming that the frozen neutrino speed of $\nu_{\mu}$ (as well its twin $\nu_{\tau}$ ) decouples from the SN $\nu_{e}$
and its antiparticle states, in a CPT conserved physics, than the question is how the  flavor states separate in flight.
Let  us notice that an Opera frozen speed  $\nu_{\mu}$ will anticipate (for Opera super-luminal neutrino velocity) a distance $\delta l_{e-\mu}=0.25 \mu$m. for each length $L$ of cm of flight.
 $$\frac{\delta l_{e-\mu}}{\frac{L}{cm}}=0.25 \cdot 10^{-4}$$
 Consequently the Compton muon neutrino wave-length $$\delta l_{\nu_{\mu_{Compton}}}=1.24 \cdot 10^{-7}\frac{10^{7} eV}{E_{\nu_{\mu}}}\mu\cdot m.$$ becomes comparable to its delayed distance (electron neutrino at light velocity)  very soon, for instance, at 10 MeV: just nearly $0.04 \mu$m. Therefore also electron anti neutrino from nuclear reactor will separate into their mass state (from muon flavors) soon depleting the $\bar{\nu_{e}}$ by a large factor, almost a half. This effect had to be observed already in atmospheric cosmic ray neutrinos and in recent years Kamland signals, see Fig \ref{fig6}. In a more remarkable way the nuclear plant energy out put would be correlated only to $57\%$ of the anti neutrino flux, contrary to well calibrated observations. Note that the so called reactor antineutrino anomaly at a few percent cannot accommodate the severe suppression above \cite{Mention}.   The atmospheric signal must combine both the early muon-electron mixing (because  super-luminal muon neutrino assumption) and the complete or partial (muon-tau) mixing. These expected de-coherence  imprint are totally absent in long known atmospheric muon and  electron neutrino anisotropy, in conflict with such ad hoc  frozen muon neutrino super-luminal speed scenario. Let  us remind that in the following  that we assume  normal 3 flavor neutrino mixing, where the probability of the muon to survive as a muon is  $P(\nu_{\mu}\rightarrow \nu_{\mu}) = 0.357$,$P(\nu_{e}\rightarrow \nu_{e}) = 0.547$,$P(\nu_{\mu}\rightarrow \nu_{e})=P(\nu_{e}\rightarrow \nu_{\mu}) = 0.264$. See Fig \ref{fig3},see Fig \ref{fig6},see Fig \ref{fig7}.

\section{Conclusions: anti-tachyon or  frozen super-luminal $\nu_{\mu}$ ?}
Assuming a nominal absolute imaginary neutrino (tachyon) mass of 117 MeV  and a Lorentz factor about 145, one may fit a tachyon signal at Opera energy and precursor time, but it is excluded because requires no SN1987A signal and a huge neutrino spread (thousand years). We imagined  a new ad hoc (possibly wrong)  anti-tachyon law (within a huge neutrino  mass about 2.4 TeV) alleviating at best this spread within 2 minutes or twelve seconds, but the model is unnatural,  with no based theoretical ground and already in remarkable conflict with  energy independence in  Opera neutrino speeds. These toy  model cannot match the well known  mixing bounds. Finally the fixed speed scenario option (where muon neutrino speed differ from electron one) also suffer of different contradictions as shown above. Finally  an even more ad hoc different frozen neutrino flavor speed, where the electron neutrino fly at (very near $(1 \mp 10^{-12})\cdot c$) speed while the  muon ones at Opera frozen super-luminal speed (very near $(1- 2.37\cdot 10^{-5})\cdot c$)) agrees (apparently)  with data,  requires a hidden SN1987A neutrino precursor in June 1983 in IMB data.  This model does suffer anyway in explaining the absent electron neutrino   mixing  within the atmospheric $\nu_{e}$ observed  behavior as well in Kamland recent records (via $\theta_{12}$ oscillation and de-coherence,see Fig \ref{fig6}, as well as the same muon neutrino depletion due to de-coherence with electron flavor in OPERA and MINOS experiment. In conclusion the imaginary neutrino mass at the needed values (for Opera-Cern claim) is in disagreement with several data and by several order of magnitude. Because of the limited time accuracy in Opera-Minos any future OPERA or MINOS  experiments, there is by present  no-go arguments, no hope to test any  observable self-consistent neutrino imaginary mass. Therefore we cannot imagine any imaginary mass able to fit the super-luminal data. Very last rumors seemed anyway to dismiss such  unbelievable discover leading to a more realistic neutrino behavior.

\section{Note after the submission}
  After  this article has been submitted, a wide sequence (hundreds) of articles in these months discussed the Opera super-luminal neutrino claim.  Earliest ones and most of all considered exotic possibilities to fit or explain the novel result \cite{ListAll}.  A few, as those   we do mention \cite{Cohen},\cite{Cowsik}  faced the eventual super-luminal consequences finding unacceptable consequences  in Cerenkov-like neutrino emission and absorption or within arguments along pion decay kinematic inconsistence leading to  a rejection of the Opera result as in our earliest and  present study. Moreover recently OPERA CERN experiment was sending much narrow bunch leading to a  confirm of their super-luminal neutrino claim \cite{Opera2}. But last minutes rumors \cite{Rumors} from OPERA seem to regard the key timing bug of the experiment. After all as someone said long time ago, Nature is subtle, but not malicious (or as we would add maliciously, a century after and later \cite{Opera2}, nor perverse). Indeed the authors thanks the same Nature that forced us to the lucky privilege  to be  defending these (now) obvious relativistic  arguments, in an embarrassing loneness, within a coral OPERA Seminar at Rome, on the 11th October 2011.

   \section{Appendix A: Neutrino mass by  Andromeda SN $\nu$ delay}
 In next nearby super-novae event, possibly from Andromeda, it would be better testable  the more conventional  time delay of the prompt neutrino masses  by their rapid neutralization NS signal versus the gravitational wave  burst \cite{Far1}. Indeed a  millisecond  prompt neutrino peak will obtain a comparable time delay (respect to SN gravitons)  due to common (real mass) neutrino slower speed, and it  may trace even the guaranteed (more mundane) real neutrino mass splitting (of atmospheric nature: $m_{\nu} \geq 0.05$ eV). In future few Mpc SN search (as toward Virgo) by future time correlated SN-GW (gravitational wave) detection the neutronization burst may lead to a neutrino mass discover. Indeed,after all, neutrino  mass may be more real than imaginary one. \\

\clearpage


\begin{thebibliography}{99}


\bibitem{Opera1} Adams T. et all. for Opera-GNGS experiment; arxiv 1109.4897v1.
\bibitem{Opera2} Adams T. et all. for Opera-GNGS experiment; arxiv 1109.4897v2.
\bibitem{Far1} D.Fargion, Lettere Al Nuovo Cimento, Volume 31, Number 15, 499-500,1981; D.Fargion, Apj. 570:909-925, 2002.
\bibitem{Far2}D.Fargion1, D. D'Armiento,P. Desiati, P. Paggi; arxiv 11012.3245
\bibitem{Ka}\emph{ http://pdg.lbl.gov/2011/reviews/rpp2011-rev-neutrino-mixing.pdf}
\bibitem{ListAll} arXiv:1109.4980, arXiv:1109.5172, arXiv:1109.5289,arXiv:1109.535,
arXiv:1109.5378,arXiv:1109.5411, arXiv:1109.5445,arXiv:1109.5599,
arXiv:1109.5671,arXiv:1109.5682,arXiv:1109.5685,arXiv:1109.5687,
arXiv:1109.5721,arXiv:1109.5727,arXiv:1109.5749,arXiv:1109.5917,
arXiv:1109.6005,arXiv:1109.6055,arXiv:1109.6097,arXiv:1109.6121,
arXiv:1109.6160,arXiv:1109.6170,arXiv:1109.6238,arXiv:1109.6249,
arXiv:1109.6282,arXiv:1109.6296,arXiv:1109.6298,arXiv:1109.6308,
arXiv:1109.6312,arXiv:1109.6520,arXiv:1109.6562,arXiv:1109.6563,
arXiv:1109.6630,arXiv:1109.6631,arXiv:1109.6641,arXiv:1109.6667,
arXiv:1109.6930,arXiv:1110.0132, arXiv:1110.0234,arXiv:1110.0239,
arXiv:1110.0241,arXiv:1110.0243,arXiv:1110.0245,arXiv:1110.0302,
arXiv:1110.0351,arXiv:1110.0392,arXiv:1110.0424,arXiv:1110.0430,
arXiv:1110.0449,arXiv:1110.0451,arXiv:1110.0456,arXiv:1110.0521,
arXiv:1110.0595,arXiv:1110.0644,arXiv:1110.0675,arXiv:1110.0697,
arXiv:1110.0736,arXiv:1110.0762,arXiv:1110.0783,arXiv:1110.0821,
arXiv:1110.0882,arXiv:1110.0931,arXiv:1110.0970,arXiv:1110.0989,
arXiv:1110.1162,arXiv:1110.1253,arXiv:1110.1317,arXiv:1110.1330.
arXiv:1110.2909,arXiv:1110.2685 ,arXiv:1110.2463,arXiv:1110.2236 ,
arXiv:1110.2219,arXiv:1110.2170, arXiv:1110.2146,arXiv:1110.2123,
arXiv:1110.2060,arXiv:1110.1943 , arXiv:1110.1875,arXiv:1110.1857,
arXiv:1110.1790.
\bibitem{Cohen} A.G.Cohen,S.Glashow; arXiv:1109.,Phys. Rev. Lett. 107, 181803 (2011)(2011)
 \bibitem{Cowsik} Cowsik R., S. Nussinov, U. Sarkar ;arXiv:1110.0241; Phys. Rev. Lett. 107, 251801 (2011).
\bibitem{Mention} G. Mention et al. .Phys. Rev. D 83, 073006 (2011)
\bibitem{Rumors} \emph{http://www.globalnews.ca/world/world/ \\european+researchers+find+flaw+in+experiment+that+measured+faster-than-light+particles/6442586764/story.html}



\end{thebibliography}
\end{document}